\documentclass[preprint,prb,aps]{revtex4}
\usepackage{amsmath,amssymb}
\usepackage{graphicx}

\begin{document}


\title{ Interdiffusion of Solvent into Glassy Polymer Films: A Molecular Dynamics Study}
\author{Mesfin Tsige\footnote{mtsige@sandia.gov}}
\author{Gary S. Grest\footnote{gsgrest@sandia.gov}}
\affiliation{Sandia National Laboratories, Albuquerque, NM 87185}
\date{\today}


\begin{abstract}
Large scale molecular dynamics and grand canonical Monte Carlo 
simulation techniques are used to study the behavior of the interdiffusion of a
solvent into an entangled polymer matrix as the state of the polymer changes 
from a melt to a glass.  The weight gain by the
polymer increases with time $t$ as $t^{1/2}$ in agreement with 
Fickian diffusion for all cases studied, although the diffusivity is found to
be strongly concentration dependent especially as one approaches the glass 
transition temperature of the polymer. The diffusivity as a
function of solvent concentration determined using the one-dimensional
Fick's model of the diffusion equation is compared to the diffusivity
calculated using the Darken equation from simulations of equilibrated 
solvent-polymer solutions. The diffusivity calculated using these two different
approaches are in good agreement. The behavior of the diffusivity strongly 
depends on
the state of the polymer and is related to the shape of the solvent
concentration profile.

\end{abstract}

\pacs{}

\maketitle

\section{Introduction}
%
Understanding the interdiffusion of solvent into a polymer is crucial for a 
variety of applications, such as food storage, controlled drug release and 
membrane separations.\cite{muller96,charati98,vegt00,yi02,lim03} The mechanisms controlling the interdiffusion process
are reasonably well understood\cite{gall89,cuthbert99,hassan99,kwan03} but predicting accurately the nature of the
diffusion has been a challenging problem. It is now widely accepted that the
interdiffusion of solvent into a polymer depends on solvent concentration
gradient within the system as well as the rate of polymer segmental relaxation.
\cite{thomas78,thomascombined,vrentas86,webb91,hopkinson97}
Whether the polymer is a melt above its glass transition temperature $T_g$ or 
an amorphous solid below $T_g$ strongly affects the diffusion behavior.\cite{thomas78}

In general, three categories of diffusion behavior of solvents into polymers
have been distinguished.\cite{sanopoulou01,greenfield01,kwan03} These are, Fickian or Case I, Case II or Class II, and
anomalous diffusion: in which the rate of diffusion of solvent is much less
than, much greater than, or comparable to the rate of polymer segmental 
relaxation, respectively. A simple descriptive way of quantifying these 
is based on the power law dependence of the mass uptake of the solvent by the
polymer or the distance covered by the solvent as a function of time $t$ ($\sim t^n$). For
Fickian diffusion $n=1/2$, for Class II diffusion $n=1$, and for
anomalous diffusion $1/2 < n < 1$.  
Fickian diffusion usually applies for all solvent concentration when the solvent
interdiffuses into a polymer melt, while
for glassy polymers it usually applies only for low solvent concentration. 
Non-Fickian kinetics is expected when the viscoelastic properties of the system
becomes the determining factor.\cite{thomas78,thomascombined}
In addition to linear kinetics, Case II diffusion is
characterized by a sharp concentration front that propagates at constant speed
\cite{thomascombined,hui87,stamatialis02}
with a Fickian type precursor foot\cite{durning95,hassan99,sanopoulou01,stamatialis02} preceding the front. 

When a solvent film is placed in contact with one surface of a polymer melt, the
diffusion is one-dimensional and can often be described by Fick's one-dimensional 
diffusion equation\cite{crank}
\begin{equation}
\frac{\partial c}{\partial t}=\frac{\partial}{\partial z}(D(c)\frac{\partial c}{\partial z}),
\label{fick}
\end{equation}
where $c$ is the solvent concentration in units of mass per unit volume and 
$D(c )$ is the diffusivity. This
equation assumes that the volume of the medium is not changed by the
interdiffusion of the solvent.
If $D(c)$ is a function of $c$ only, then the Boltzmann transformation of 
Eq.~\ref{fick} gives
\begin{equation}
\frac{\partial z}{\partial t}=f(D,c)t^{1/2}
\label{sqt}
\end{equation}
where $f(D,c)$ is a function of $D$ and $c$ only. This equation reflects the 
square root time dependence of Fickian diffusion irrespective of the functional
form of $D(c)$. It can be integrated to yield the diffusion coefficient at 
concentration $c^{\prime}$ \cite{crank}
\begin{equation}
D(c^{\prime})=-\frac{1}{2}\left[\left(\frac{dc}{d\eta}\right)\right]_{c^{\prime}}^{-1} \
\int_{0}^{c^{\prime}}\eta dc
\label{diffusivity}
\end{equation}
where $\eta=z/t^{1/2}$. Thus from the scaled concentration profile one can
directly obtain the diffusivity $D(c)$. Note that only in a few special cases,
like $D(c)$ constant, can Eq.~\ref{fick} be solved analytically.\cite{crank} 
For the case in which $D(c)$ is a constant, $c(\eta)$ is an error function.
However, it is important to remember that Fickian diffusion, i.e. uptake 
increasing as $t^{1/2}$, is true
provided that the diffusivity depends only on $c$ as shown above.

The diffusivity can also be
approximately obtained from the following Darken equation applied to solvent 
diffusing in an equilibrated polymer solution
\begin{equation}
D(c)=D_c(c)\left(\frac{\partial\ln f}{\partial\ln c}\right)_T
\label{darken}
\end{equation} 
where $D_c(c)$ is the corrected diffusion constant and $f$ is the fugacity of 
the solvent, both are defined in the next section.

Molecular dynamics (MD) simulation technique is proven to be a
useful tool for determining the diffusion coefficients of penetrant molecules
in polymers. This technique is specially important when detailed microscopic
information of the mechanism of transport is required. Most of the previous 
studies have focused on the penetrant transport of small molecules in a polymer
melt.
\cite{sonnenburg90,takeuchi90,pant92,muller92,sok92,tamai94,gee95,hofman96,li97,muller98,hahn99,vegt00,lim03}
With recent advances in parallel molecular dynamics algorithms and the increased
computational power, progress has primarily occurred for studying the diffusion
of large molecules (phenol molecules) in a polymeric matrix at atomistic 
level.\cite{hahn99,karlsson02}
However, equivalent development is lacking for interdiffusion of solvent into 
polymer or polymer-polymer interdiffusion. In the previous study, which will 
be referred to hereafter as paper I,\cite{mesfin04} we investigated the 
interdiffusion of a solvent into a homopolymer melt. The solvent concentration
profile and weight gain by the polymer was measured as a function of time. The
weight gain was found to scale as $t^{1/2}$ and the concentration profiles
were found to fit very well assuming Ficks's second law with constant 
diffusivity. The study, however, focused only on homopolymers that are far above
the glass transition temperature.

In this paper we extend our previous study on interdiffusion of solvent into
homopolymers that are close to the glass transition temperature of
the homopolymer. 
Case II diffusion has been inaccessible to computer simulation due to the 
extensive computational effort required. In this paper we study the conditions 
that may lead to Case II type diffusion behavior.
The main purpose of the present study is two-fold: first, to understand how the
dependence of the diffusivity $D(c)$ on concentration changes as the state
of the polymer changes and also its relation to the form of the concentration
profile curve; second, to test the Darken approach under general conditions
where $D_c(c)$ is not a constant. As in paper I, we are also interested in the
relationship between the self-diffusion constant of the solvent and the 
corrected diffusion and the diffusivity.

The outline of this paper is as follows. In Sec. II a brief review of the molecular
dynamics simulation and the model used is given. In Sec. III the interdiffusion
results for different polymer-polymer and solvent-polymer interaction parameters
are presented and discussed. The diffusivity $D(c)$ calculated from solvent
concentration profiles and from the Darken equation are presented in Sec. IV. 
The self- and corrected diffusion constants as a function of
solvent concentration are also presented and discussed. The main results of the
present study are summarized in Sec. V.

\section{Simulation Details}

\subsection{System}
The basic model of the polymer-solvent system is the same as used in paper I. 
The polymer is treated as freely jointed bead-spring chain of length $N$
monomers of mass $m$ and the solvent is modeled as single monomer of mass $m$. 
All monomers of type $\alpha$ and $\beta$ interacts through the standard
Lennard-Jones 6-12 potential
\begin{equation}
U_{LJ}(r)=\left\{ \begin{array}{ll}
4\epsilon_{\alpha\beta}\left\{\left(\frac{\sigma_{\alpha\beta}}{r}\right)^{12}-\left(\frac{\sigma_{\alpha\beta}}{r}\right)^{6}\right\}+\epsilon_{LJ}, & r\le r_c\\
0, & r>r_c
\end{array} \right.
\end{equation}
where $r$ is the distance between monomers and $\epsilon_{LJ}$ is a constant
added so that the potential is continuous at $r=r_c$.  Here $\alpha=p$ stands 
for the polymer monomer and $\alpha=s$ for a solvent monomer.  
$\epsilon_{ss}=\epsilon$ and $\sigma$ define the units of energy and length, 
respectively. Here we take $\sigma=\sigma_{\alpha\beta}$ and $r_c=2.5\sigma$.
For our model, the freezing temperature of the solvent is higher than the glass
transition temperature of a long fully flexible polymer melt, 
$T_g=0.5-0.6\epsilon/k_B$.\cite{baljon01} Thus, temperature is not a good 
variable to change
the state of the polymer without changing the state of the solvent. 
Instead, we vary $\epsilon_{pp}$ from $\epsilon_{pp}=\epsilon$ (melt) to
$2\epsilon$ (glassy). Berthelot rule 
$\epsilon_{sp}=\sqrt{\epsilon_{ss}\epsilon_{pp}}$ is used for the cross term
in some cases but we also study other cross terms. This is because
($\epsilon_{pp},\epsilon_{sp})=(2\epsilon, \sqrt{2}\epsilon)$ is found
to be immiscible except in the dilute limit. As in paper I, for bonded
monomers an additional anharmonic potential known as FENE potential with 
$R_0=1.5\sigma$ and $k=30\epsilon$ is applied.\cite{grest86,kremer90}

For the interdiffusion of solvent monomers into a polymeric matrix, the
system consist of entangled polymer chains in a rectangular box which is periodic in $x$ and $y$ but not in
$z$, the diffusion direction. This initial configuration was generated following the
procedure given in paper I. The polymer consisted of 600
chains of length $N=500$ monomers. The solvent consisted of~230,000 monomers. 
For the self- and corrected diffusion studies as a function of concentration,
the system consist of an equilibrated polymer solvent mixture in a cubic
box which is periodic in all directions. The polymer in the system consisted of
$M$ chains of length $N=500$ monomers, where $M=100$ for solvent concentration
$c < 0.45\sigma^{-3}$ and $M=50$ for $c \ge 0.45\sigma^{-3}$. The mole fraction
of solvent $x_s$ in the mixture was varied from 0.01 (dilute case) to 0.75.
A pure solvent system of 50,000 monomers was also simulated.

In paper I we compared results from Langevin thermostat simulations which 
screens the hydrodynamic interactions with results from dissipative particle
dynamics (DPD) thermostat simulations, which does not. The results from the
two thermostats agreed when the dissipation from the thermostats become much
smaller than from particle collisions. In the present study, to conserve
hydrodynamic interactions, we use DPD thermostat through out the simulation. 
For details see paper I.
The equations of motion were integrated with a velocity verlet algorithm 
with a time step of $\Delta t=0.012\tau$ for the interdiffusion study and
$\Delta t=0.009\tau$ for the bulk equilibrium measurement of the self- and
corrected diffusion constants, where $\tau=m(\sigma/\epsilon)^{1/2}$.
All the simulations were run  using the massively
parallel code LAMMPS\cite{steve} at a temperature of $T=\epsilon/k_B$ and 
pressure $P\simeq0$ without tail corrections to be comparable to interdiffusion
-- same as paper I.

\subsection{Diffusion Coefficients}

We have calculated the self-and corrected diffusion coefficients and diffusivity
$D(c)$ of solvents in an equilibrated solvent polymer mixture as a function
of solvent concentration, $c$. 
The self-diffusion constant $D_s(c)$ of the solvent in the polymer was
calculated from the slope of the solvent mean square displacement
\begin{equation}
D_s(c)=\lim_{t \rightarrow \infty}\frac{\langle[\boldsymbol{r}(t)-\boldsymbol{r}(0)]^2\rangle}{6t}.
\label{tracer}
\end{equation}
Here $\langle...\rangle$ denote an ensemble average
and is obtained by averaging over all solvents and many initial time origins,
and $\boldsymbol{r}(t)$ is the position of a solvent in the polymer at time $t$.

The corrected diffusion coefficient $D_c(c)$ of the solvent in the polymer is
calculated using the Einstein form equation\cite{mesfin04}
\begin{eqnarray}
D_c(c) &=& N_Tx_sx_p\lim_{t\rightarrow \infty}\frac{1}{6t}\langle \{\left[\boldsymbol{r}_{cm,s}(t)-\boldsymbol{r}_{cm,s}(0)\right]   \nonumber \\
& & -\left[\boldsymbol{r}_{cm,p}(t)-\boldsymbol{r}_{cm,p}(0)\right]\}^2\rangle
\label{einstein}
\end{eqnarray}
where $x_i$ and $\boldsymbol{r}_{cm,i}(t)$ are mole fraction and center of mass
of all monomers of species $i$ at time $t$, respectively, and $N_T=N_s+N_p$ is 
the total number of monomers.

To determine the fugacity $f$ the particle insertion 
method\cite{heffelfinger98} is applied using the grand canonical 
MD code LADERA.\cite{aidan} During the course of an equilibrium molecular dynamics 
simulation at the appropriate solvent concentration, the energy, $E$, of 
inserting a solvent particle at random locations was sampled. The excess
chemical potential energy $\mu_e$ is computed using
\begin{equation}
\mu_e=-k_BT\ln\langle\exp(-E/k_BT)\rangle,
\end{equation}
where $k_B$ is the Boltzmann constant, $T$ is the temperature and 
$\langle...\rangle$ is an ensemble average. Then, the activity coefficient, 
$\gamma$, is computed via $\gamma=\exp(\mu_e/k_BT)$.
The thermodynamic factor in eq.~\ref{darken} can be expressed in
terms of the activity coefficient $\gamma$ of the solvent as
\begin{equation}
\frac{\partial \ln f}{\partial \ln c}=1+\frac{\partial \ln \gamma}{\partial \ln c}.
\label{thermodynamic}
\end{equation}
The thermodynamic factor goes to 1 as $c\rightarrow 0$.

The computing time depends on the state of the polymer where much longer run
are required as the effective temperature of the polymer melt is reduced towards
its glass transition temperature. At the lowest temperature studied and for
a given solvent concentration $c$ a run of
about half a million MD time steps are required to calculate the self-diffusion
constant $D_s(c)$ while a run of more than four million MD time steps is 
required for the corrected diffusion constant $D_c(c)$. The fugacity calculation
at each solvent concentration requires more than four million MC insertion 
attempts. For the interdiffusion simulations, one run was made for each 
set of parameters specified. Each system is run until the solvent reaches
the lower substrate and on the average requires a run of about
five million MD time steps.

\section{Interdiffusion}

\begin{figure}[bth!]
\begin{center}
\includegraphics*[width=3.3in]{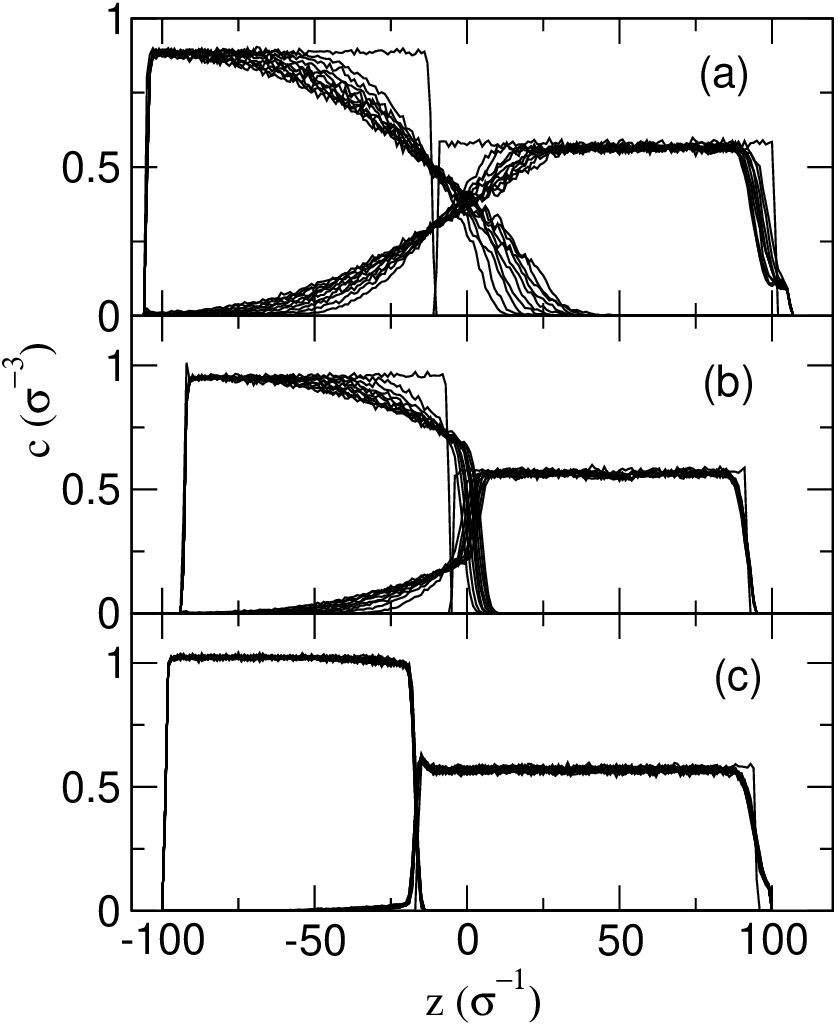}
\end{center}
 \caption{
 Solvent and polymer concentration profiles as a function
 of time for (a) $k_BT/\epsilon_{pp}=1.0$, plotted every 2400$\tau$,
 (b) $k_BT/\epsilon_{pp}=0.75$, plotted every 4800$\tau$, and (c) 
 $k_BT/\epsilon_{pp}=0.5$,
 plotted every12000$\tau$, with Berthelot's rule $\epsilon_{sp}=
 \sqrt{\epsilon_{pp}\epsilon_{ss}}$ for the cross-term. The solvent
 diffusing into the polymer from the right side.
 }
 \label{fig:berthelot}
 \end{figure}

\begin{figure}[bth!]
\begin{center}
\includegraphics*[width=3.0in]{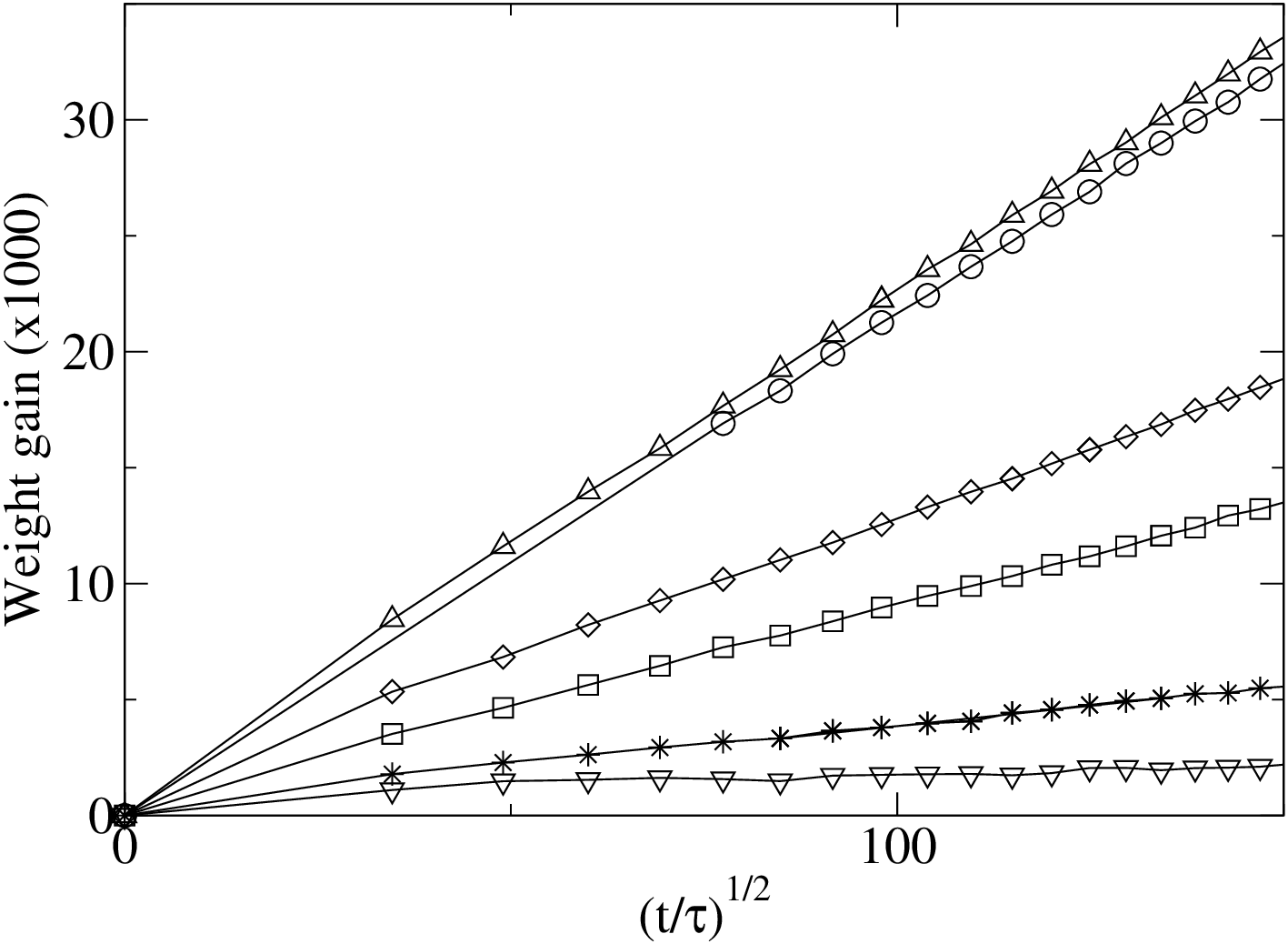}
\end{center}
 \caption{
Weight gain in terms of the number of solvent monomers diffused into the 
polymer at a given time $t$, for $(\epsilon_{pp}$, $\epsilon_{sp})$=
$(\epsilon$, $\epsilon)$ (circles), $(1.33\epsilon$, $\sqrt{1.33}\epsilon)$ 
(squares), $(2.0\epsilon$, $\sqrt{2.0}\epsilon)$ (down triangles), 
$(2.0\epsilon$, $1.55\epsilon)$ (stars), 
$(2.0\epsilon$, $1.7\epsilon)$ (diamond), and 
$(2.0\epsilon$, $2.0\epsilon)$ (up triangles), $T=\epsilon/k_B$.
 }
 \label{fig:weightgain}
 \end{figure}

Interdiffusion studies of a solvent into an equilibrated polymer has been 
conducted for different cases of polymer-polymer and solvent-polymer interactions. 
The initial setup for the interdiffusion study is the same as Fig. 1 of paper I.
The density profile
of both polymer and solvent as a function of time for 
$k_BT/\epsilon_{pp}=1.0$, 0.75 and 0.5 with Berthelot's rule for the
cross-term $\epsilon_{sp}$ is shown in Fig.~\ref{fig:berthelot}. The solvent
diffuses into the
polymer from the right side. As the state of the polymer changes from melt
($k_BT/\epsilon_{pp}=1.0$) to glass ($k_BT/\epsilon_{pp}=0.5$) the solvent
density profile changes to a sharp front. The solvent diffusion is Fickian in 
all the three cases as confirmed by the
linearity of the weight gain by the polymer versus $t^{1/2}$ curve shown in
Fig.~\ref{fig:weightgain}. This indicates that the precursor of the front for 
$k_BT/\epsilon_{pp}=0.5$ is Fickian. This is in agreement with recent 
experimental observations that characterize Case II diffusion by a sharp 
concentration front with a Fickian type precursor.\cite{durning95,hassan99,sanopoulou01,stamatialis02} 
For the front to move the solvent mobility should be much greater than the 
rate of polymer segmental relaxation.\cite{thomas78} But, for this case the 
front does not move in the time scale of our simulation. In fact, 
simulation of an equilibrated solvent-polymer solution, discussed in the next 
section, shows that only a small amount of solvent is soluble for this case 
suggesting that the front may not move at all.

\begin{figure}[bth!]
\begin{center}
\includegraphics*[width=3.3in]{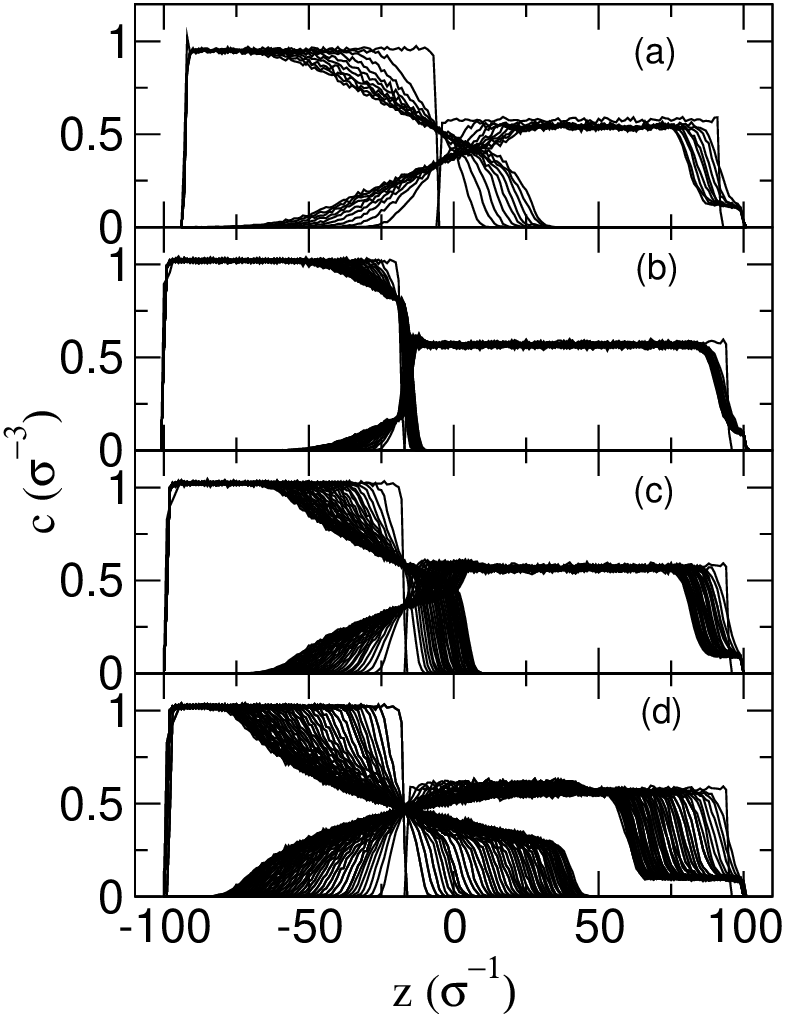}
\end{center}
 \caption{
Solvent $\rho_s$ and polymer $\rho_p$ concentration profiles as a function
of time, plotted every 2400$\tau$ for ($\epsilon_{pp},\epsilon_{sp})=$
(a) (1.33$\epsilon$,1.33$\epsilon$), (b) (2.0$\epsilon$,1.55$\epsilon$),
(c) (2.0$\epsilon$,1.7$\epsilon$), and (d) (2.0$\epsilon$,2.0$\epsilon$).
 }
 \label{fig:e22}
 \end{figure}

In order to facilitate the interdiffusion of solvent into the glassy polymer
($k_BT/\epsilon_{pp}=0.5$), the interaction between polymer and solvent is
increased to $\epsilon_{sp}=1.55\epsilon$, $1.7\epsilon$, and $2.0\epsilon$.
Similarly $\epsilon_{sp}$ is increased to 1.33$\epsilon$ for 
$k_BT/\epsilon_{pp}=0.75$.
The corresponding density profiles of both polymer and solvent as a function of
time is shown in Fig.~\ref{fig:e22}. We clearly see that as $\epsilon_{sp}$
is increased the solubility is enhanced for both $\epsilon_{pp}=1.33\epsilon$ 
and $2.0\epsilon$ and the density front observed for 
($\epsilon_{pp},\epsilon_{sp})=(2.0\epsilon,\sqrt{2.0}\epsilon)$ disappears. 
The $\epsilon_{sp}=1.7\epsilon$ case shows a cross-over. There is no
change in the diffusion process due to the change in $\epsilon_{sp}$ as the
weight
gain by the polymer system for all four cases increases as $t^{1/2}$, see Fig.~\ref{fig:weightgain}, in
agreement with Fickian diffusion.

\section{Diffusion Coefficients}

The diffusivity of the solvent can be calculated from the solvent concentration
profile using Eq.~\ref{diffusivity}. It can be also calculated from simulation of
equilibrated solvent-polymer solution using the Darken equation (Eq.~\ref{darken}).
In this section we compare diffusivity results from the two different approaches.

\subsection{Diffusivity from concentration profile}

\begin{figure}[bth!]
\begin{center}
\includegraphics*[width=3.3in]{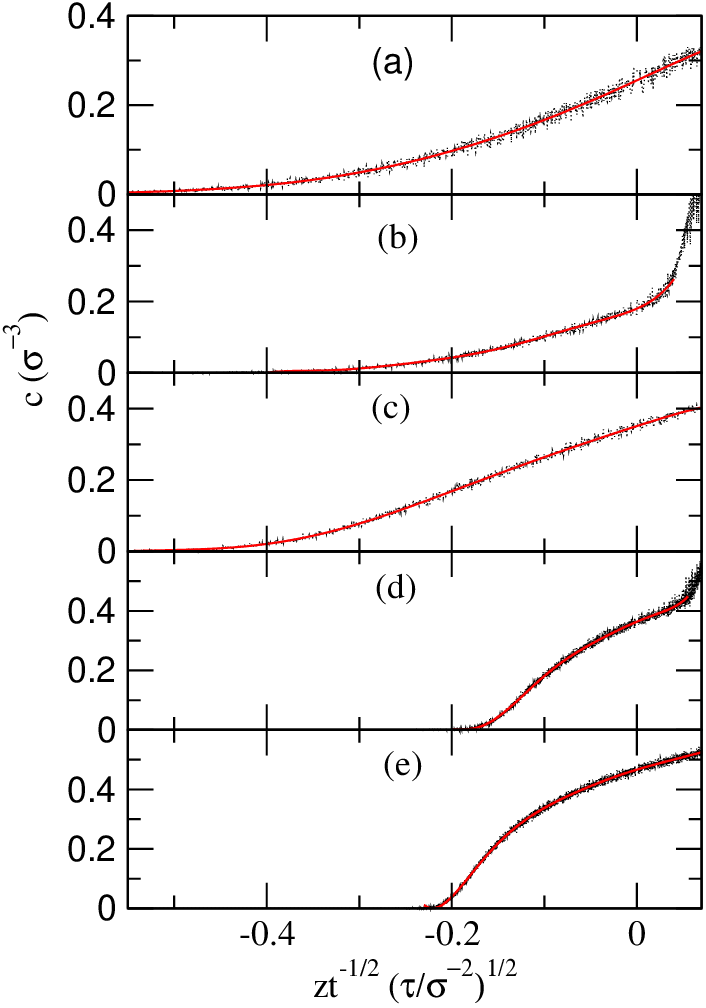}
\end{center}
 \caption{
 [Color online] Solvent concentration profiles are plotted as a function of the scaling 
 variable $zt^{-1/2}$ for ($\epsilon_{pp},\epsilon_{sp})=$ (a)
 ($\epsilon$, $\epsilon$), (b) (1.33$\epsilon$, $\sqrt{1.33}\epsilon$),
(c) (1.33$\epsilon$, 1.33$\epsilon$), (d) (2.0$\epsilon$,1.7$\epsilon$),
and (e) (2.0$\epsilon$,2.0$\epsilon$). The red (light dark) solid lines represent
the theoretical curve based on the solution of Eq.~\ref{diffusivity}.
 }
 \label{fig:normalized}
 \end{figure}

Using the change of variable $\eta=zt^{-1/2}$ the solvent density profiles
corresponding to different times superimpose as shown
in Fig.~\ref{fig:normalized}. This indicates that the solvent diffusivity
is independent of position as expected for Fickian diffusion. It also means that
the solvent diffusivity for absorption can be determined as a function of
concentration by integrating Eq.~\ref{diffusivity} with respect to $\eta$.

\begin{figure}[bth!]
\begin{center}
\includegraphics*[width=3.0in]{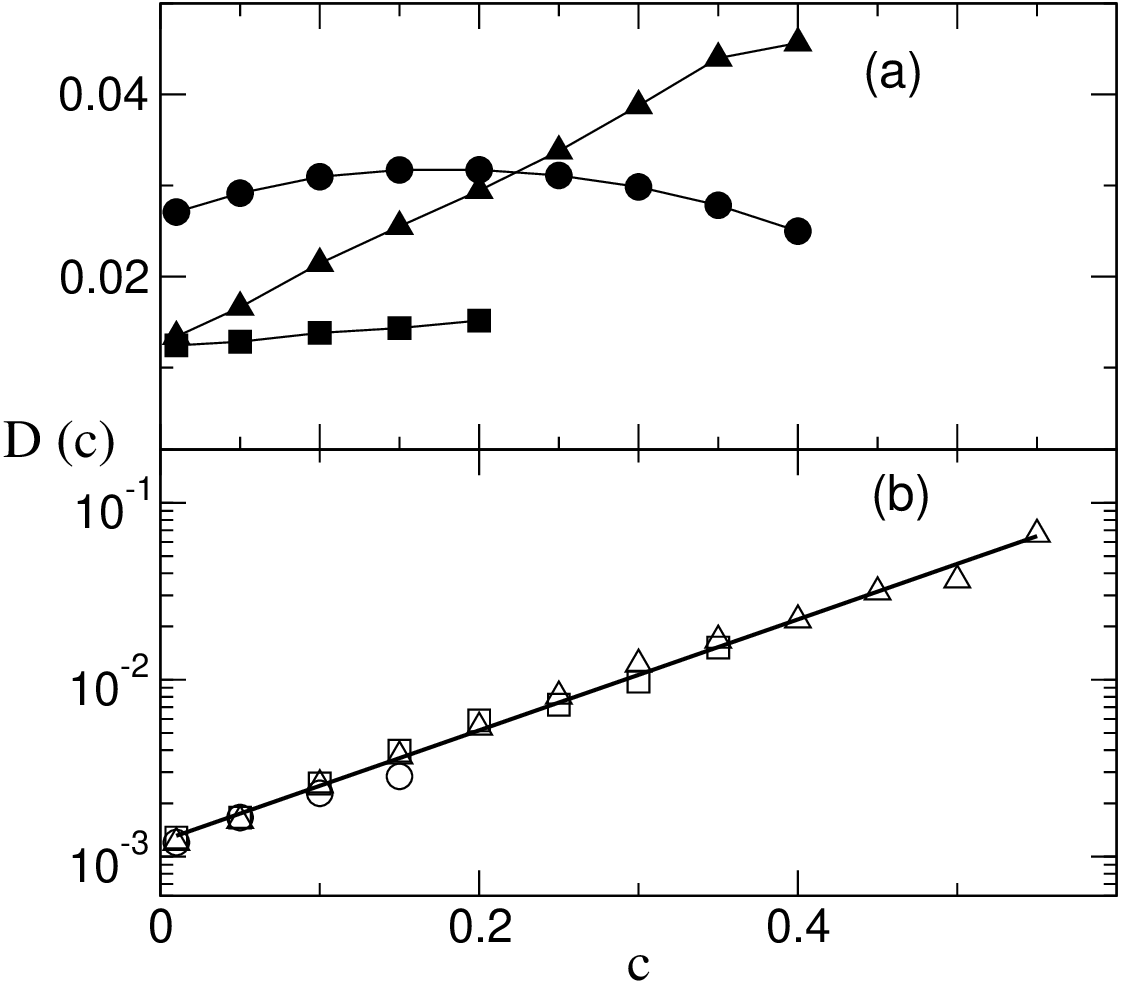}
\end{center}
 \caption{
Diffusivity $D(c)$ as a function of solvent concentration calculated from the solvent
concentration profile using Eq.~\ref{diffusivity}. The different symbols are for
$(\epsilon_{pp}$, $\epsilon_{sp})=$ $(\epsilon$, $\epsilon)$ (closed
circles), $(1.33\epsilon$, $\sqrt{1.33}\epsilon)$ (closed squares),
$(1.33\epsilon$, $1.33\epsilon)$ (closed triangles), 
$(2.0\epsilon$, $1.55\epsilon)$ (open circles), 
$(2.0\epsilon$, $1.7\epsilon)$ (open squares), and
$(2.0\epsilon$, $2.0\epsilon)$ (open triangles). The solid line in (b) is an
exponential fit to the data.
}
\label{fig:calculated}
\end{figure}

To calculate $D(c)$ analytically using Eq.~\ref{diffusivity}, the average of
the transformed solvent density profiles $c(\eta)$ of each case shown in 
Fig.~\ref{fig:normalized}
was fit to a polynomial function of at least order 5. This higher order 
polynomial was chosen to get an optimal fit to the concentration profiles. 
The data is integrated analytically up to the target concentration using the
transformation $\int_{0}^{c^{\prime}}\eta dc=\int_{\eta_0}^{\eta}\eta\frac{dc}{d\eta} d\eta$.
This procedure is repeated for different values of solvent 
concentration and the diffusivity $D(c)$ calculated for the 
different cases is shown in Fig.~\ref{fig:calculated}.

In general, the behavior of $D(c)$ strongly depends on the state of the polymer.
The diffusivity is approximately a constant when the homopolymer is far above 
the glass
transition, $T_g$, that is for $k_BT/\epsilon_{pp}=1.0$, and then becomes 
concentration dependent as the effective temperature of the polymer melt
$k_BT/\epsilon_{pp}$ is reduced towards its glass transition temperature. For
($\epsilon_{pp},\epsilon_{sp})=(1.33\epsilon,\sqrt{1.33}\epsilon)$,
$(2.0\epsilon, \sqrt{2.0}\epsilon)$ and $(2.0\epsilon, 1.7\epsilon)$ the 
calculated diffusivity for 
concentrations in the steep part of the concentration profile (not included in 
Fig.~\ref{fig:calculated}) resulted in large scatter of the data. This is
because little variation of the slope in this region introduces large error
in the diffusivity. 

For $\epsilon_{pp}=1.33\epsilon$, the diffusivity is approximately a constant for
$\epsilon_{sp}=\sqrt{1.33}\epsilon$, but increases linearly with concentration
for $\epsilon_{sp}= 1.33\epsilon$. Note that in the dilute limit ($c\rightarrow0$)
the diffusivity is independent of $\epsilon_{sp}$. For 
$\epsilon_{pp}=2.0\epsilon$, the diffusivity is independent of $\epsilon_{sp}$
within the error of the simulation.  For this case the diffusivity can be 
approximated by an exponential function of the form $D(c)=D_0\exp(\alpha c)$,
where $D_0$ and $\alpha$ are constants that may depend on the state of the
polymer, and is represented by a line in 
Fig.~\ref{fig:calculated}(b). The empirical expression found from
fitting the diffusivity curve of a given system was in turn used to solve 
Eq.~\ref{fick} and the 
calculated concentration profiles are shown in Fig.~\ref{fig:normalized} as 
solid lines. In all cases, the calculated concentration
profiles give an adequate description of the simulated concentration profiles
in the region of interest.

The behavior of $D(c)$ is directly related to the form of the concentration
profile curve. When the solvent concentration profile is concave 
(i.e. for ($\epsilon_{pp},\epsilon_{sp})=(\epsilon, \epsilon)$ and 
$(1.33\epsilon, \sqrt{1.33}\epsilon)$) the diffusivity is approximately a constant. 
However, when the 
diffusivity shows exponential dependence on solvent concentration (for
$\epsilon_{pp}=2.0\epsilon$) then the solvent concentration profile curve 
becomes convex. For the case in which the concentration is approximately linear 
($\epsilon_{pp}=\epsilon_{sp}=1.33\epsilon$), the diffusivity increases
linearly with concentration. Numerical solution of Eq.~\ref{fick} for a
given functional form of diffusivity on concentration has shown a similar 
relation between the shape of the concentration profile and the
dependence of diffusivity on concentration.\cite{crank}

\subsection{Diffusion Coefficients from Equilibrated Polymer solution simulations}

\begin{figure}[bth!]
\begin{center}
\includegraphics*[width=3.0in]{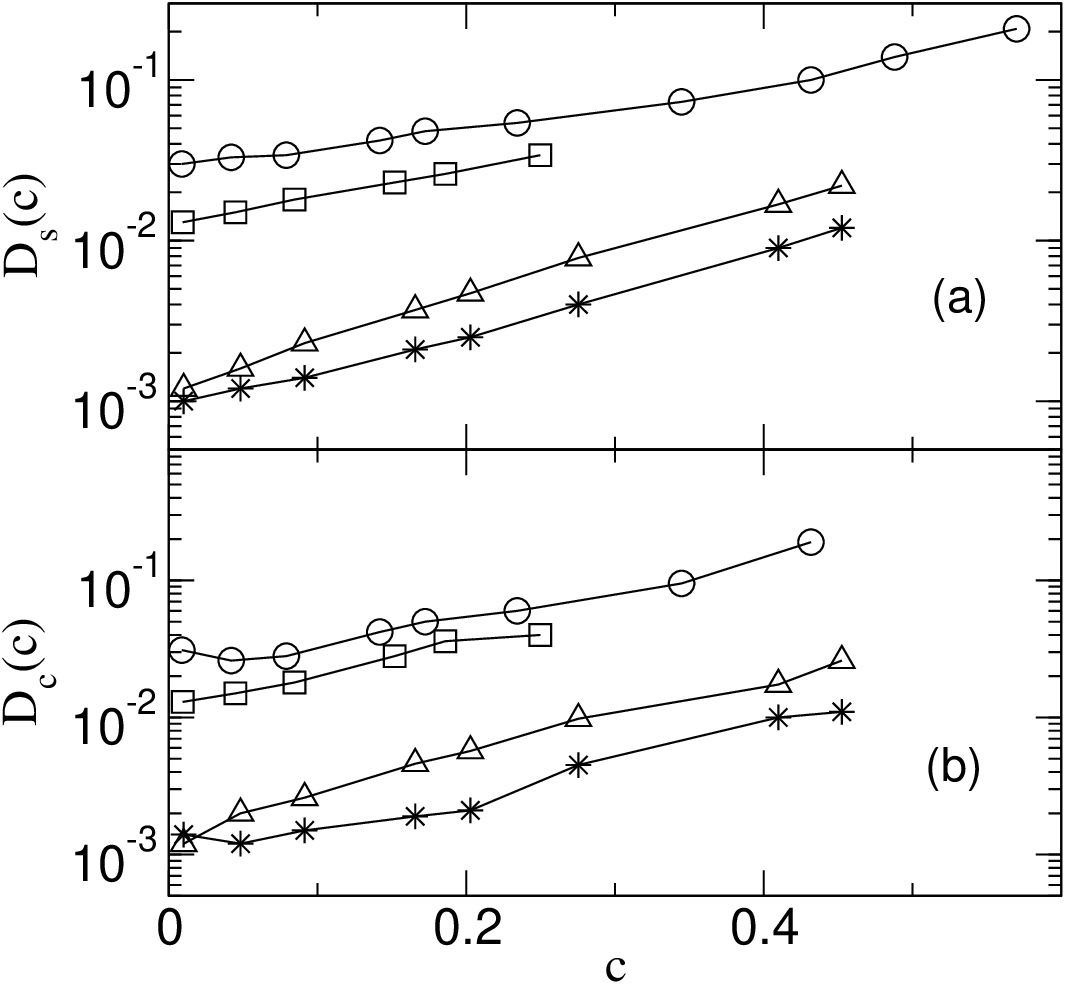}
\end{center}
 \caption{
Dependence of diffusion constants on solvent concentration, (a) $D_s(c)$ and
(b) $D_c(c)$. Symbols are for $(\epsilon_{pp}$, $\epsilon_{sp})=$
$(\epsilon$, $\epsilon)$ (circles), $(1.33\epsilon$, $\sqrt{1.33}\epsilon)$
(squares), $(2.0\epsilon$, 1.7$\epsilon)$ (triangles), and 
$(2.0\epsilon$, 2.0$\epsilon)$ (stars).
 }
 \label{fig:self}
 \end{figure}

{\bf Self- and corrected diffusion coefficients.}
The self-diffusion, $D_s(c)$, and corrected diffusion, $D_c(c)$, coefficients
are calculated from Eq.~\ref{tracer} and Eq.~\ref{einstein}, respectively. $D_s(c)$
and $D_c(c)$ as a function of solvent concentration for different 
polymer-polymer and solvent-polymer interactions are shown in 
Fig.~\ref{fig:self}(a) and (b), respectively. Note that the diffusion
coefficients calculated for ($\epsilon_{pp},\epsilon_{sp})=(1.33\epsilon,
\sqrt{1.33}\epsilon)$ and $(2.0\epsilon, \sqrt{2.0}\epsilon)$ is limited to low 
solvent concentration since the systems
phase separate for large $c$. The critical value
of solvent concentration above which the system phase separates can be 
approximately
determined directly from the behavior of the solvent concentration profile shown in
Fig.~\ref{fig:berthelot} and \ref{fig:e22}. We have observed that for
($\epsilon_{pp},\epsilon_{sp})=(1.33\epsilon, \sqrt{1.33}\epsilon),
(2.0\epsilon,\sqrt{2.0}\epsilon)$, and $(2.0\epsilon, 1.7\epsilon)$ the 
system phase separates for concentration values corresponding to the steep part
of the concentration profile. The critical solvent concentration value is
approximately the concentration value corresponding to the inflection point of the
concentration curve. The critical solvent concentration value for 
($\epsilon_{pp},\epsilon_{sp})= (2.0\epsilon,\sqrt{2.0}\epsilon)$ is basically 
the dilute limit. 

In general, the self and corrected diffusion coefficients shown in 
Fig.~\ref{fig:self} show an exponential dependence on concentration. However,
for ($\epsilon_{pp},\epsilon_{sp})=(\epsilon, \epsilon)$ and
$(1.33\epsilon, \sqrt{1.33}\epsilon)$
the dependence of the corrected diffusion on concentration is weak at low solvent 
concentration. For a given value of concentration, as expected, both $D_s(c)$ 
and $D_c(c)$ decrease as the state of the polymer changes from melt to glassy.
Note that $D_s(0)\simeq D_c(0)$ in all cases.

\begin{figure}[bth!]
\begin{center}
\includegraphics*[width=3.0in]{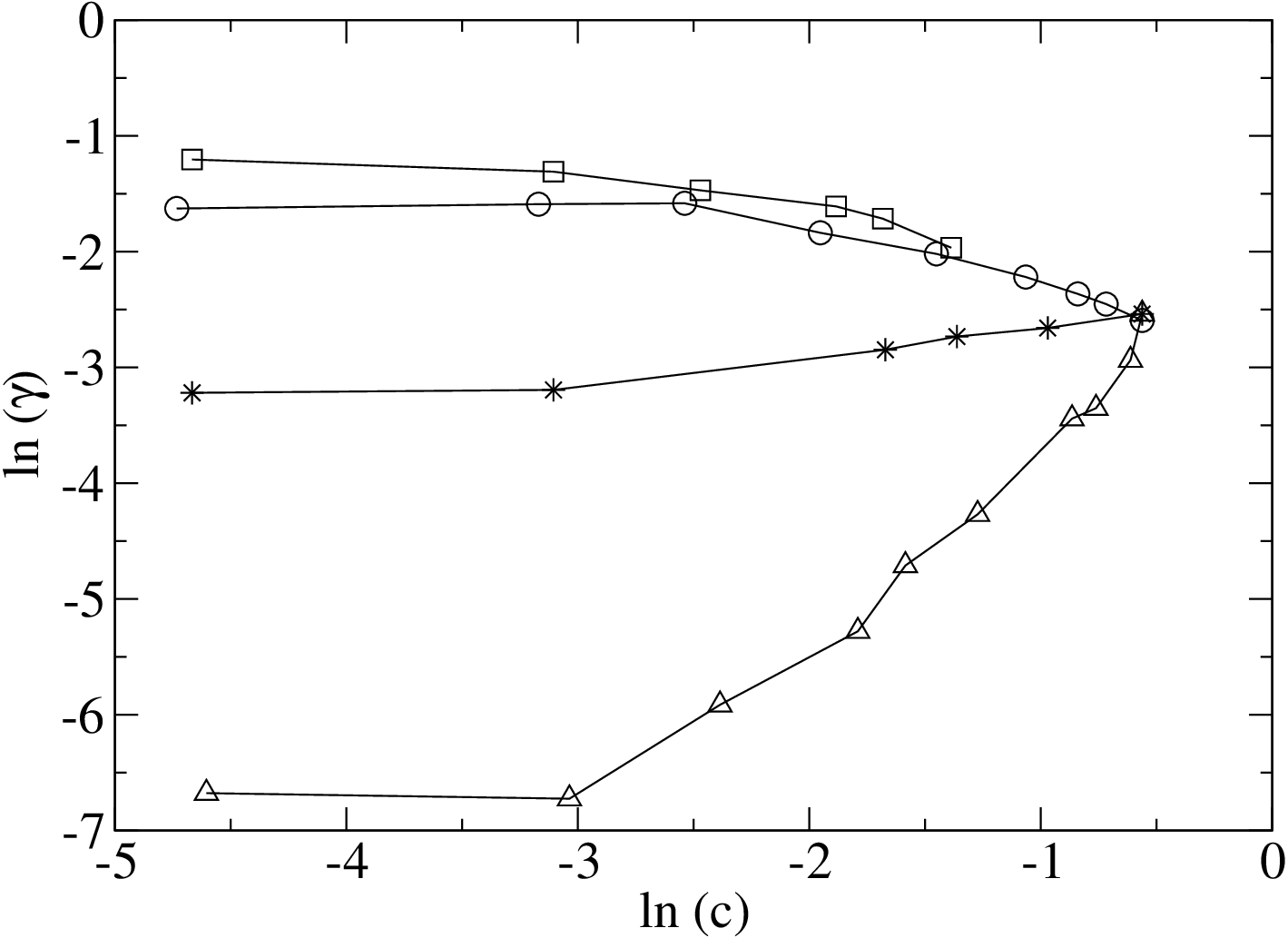}
\end{center}
 \caption{
Activity coefficient of solvent as a function of solvent concentration.
Symbols are for $(\epsilon_{pp}$, $\epsilon_{sp})=$
$(\epsilon$, $\epsilon)$ (circles), $(1.33\epsilon$, $\sqrt{1.33}\epsilon)$
(squares), $(2.0\epsilon$, 2.0$\epsilon)$ (triangles), and 
$(1.33\epsilon$, 1.33$\epsilon)$ (stars)
}
\label{fig:activity}
\end{figure}

{\bf Diffusivity.}
To calculate the diffusivity $D(c)$ using Eq.~\ref{darken}, the thermodynamic
factor given by Eq.~\ref{thermodynamic} has to be first determined. Using the
GCMD simulation method, the activity coefficient of the solvent is determined
as a function of solvent concentration and is shown in Fig.~\ref{fig:activity} 
on a ln-ln plot. Note that the concentration at which all the activity 
coefficients converge is the pure solvent case.
As expected, the activity coefficient is constant for low solvent concentration
and thus $D(0)\approx D_c(0)=D_s(0)$ for all cases. As the solvent concentration
increases the activity
coefficient for ($\epsilon_{pp},\epsilon_{sp})=(\epsilon, \epsilon)$ and
$(1.33\epsilon, \sqrt{1.33}\epsilon)$ decreases with solvent concentration and
increases for $\epsilon_{pp}=\epsilon_{sp}=1.33\epsilon$ and 
$2.0\epsilon$. This results in the diffusivity for the former two cases to be
lower while for the latter two cases to be higher than the corrected diffusion 
constant. However, the curves in Fig.~\ref{fig:activity} are not smooth, making
it difficult to determine the thermodynamic factor and thus the diffusivity with
high precession.

\begin{figure}[bth!]
\begin{center}
\includegraphics*[width=3.0in]{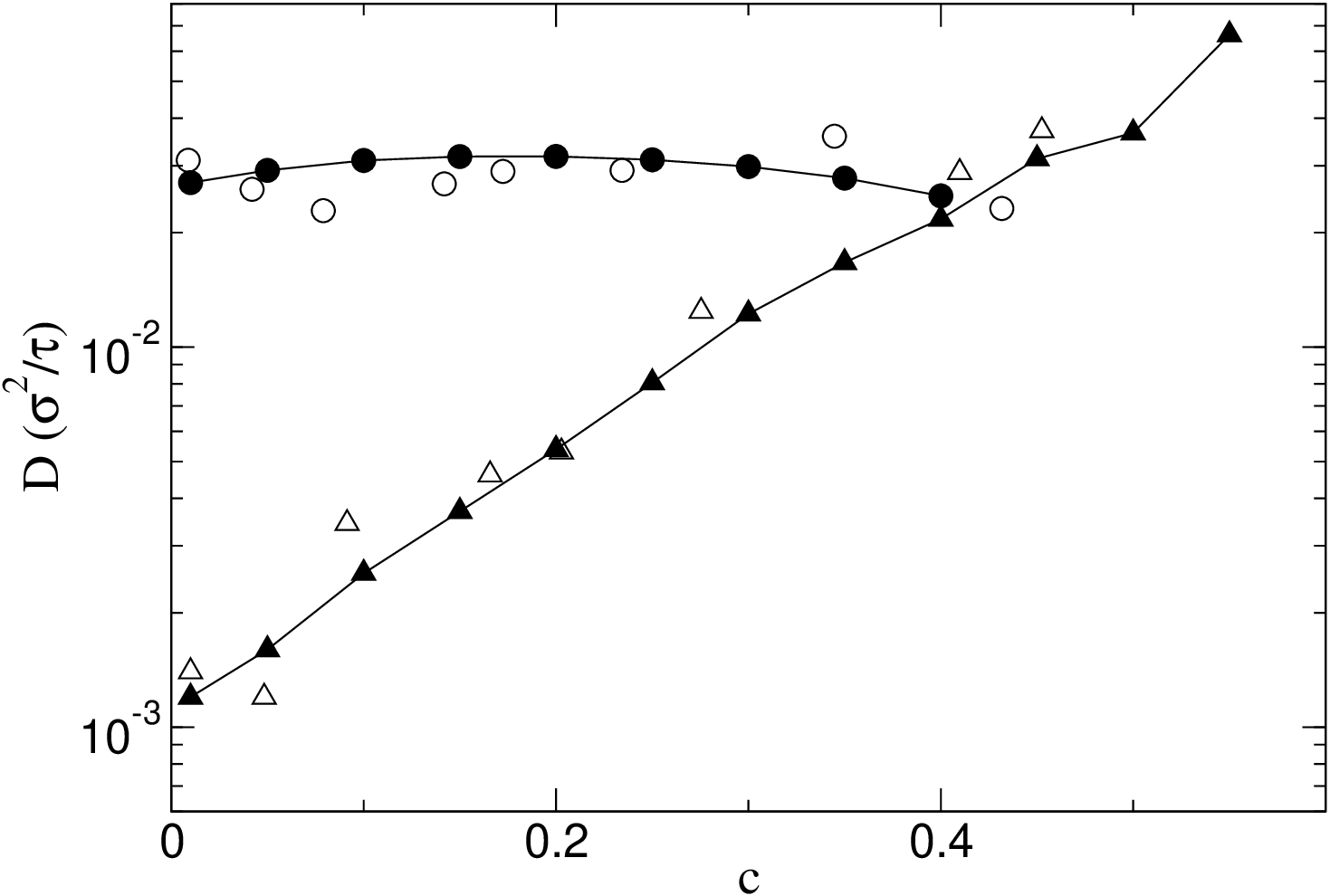}
\end{center}
 \caption{
Diffusivity $D(c)$ as a function of solvent concentration and open symbols are from
Darken equation Eq.~\ref{darken} and closed symbols are from the concentration
profile using Eq.~\ref{diffusivity}. Circles and triangles are for
$\epsilon_{pp}=\epsilon_{sp}=\epsilon$, and $2.0\epsilon$, respectively.
}
\label{fig:diffusivity}
\end{figure}

Instead of numerically differentiating the activity curves as we did in
paper I, the activity curve for each case was fit to a cubic function and the
thermodynamic factor was determined analytically.  The diffusivity calculated
following this procedure is shown in Fig.~\ref{fig:diffusivity}.
For comparison, the diffusivity calculated from the solvent concentration 
profile reported in Sec. IVA is also shown as closed symbols. In general, the
diffusivity calculated from the two different approaches show good agreement.
The diffusivity calculated using the Darken equation shows the expected
behavior discussed above, the diffusivity is constant for
$\epsilon_{pp}=\epsilon$ and increases exponentially
for $\epsilon_{pp}=\epsilon_{sp}=2.0 \epsilon$. However, there is more
scatter due to the uncertainty in the thermodynamic factor.

In the interdiffusion study the behavior of $D(c)$ is related to the shape of 
the solvent concentration profile. Similarly, in the present approach the 
behavior of $D(c)$ can be directly related to the functional form of the 
activity coefficient. In general, for a constant diffusivity the activity 
coefficient decreases with concentration, and increases with concentration for
a diffusivity that increases with concentration.

\section{Summary}

The effect of polymer-polymer and solvent-polymer interactions on the behavior of
the interdiffusion of a solvent in to an entangled polymer matrix have been 
studied
using large scale molecular dynamics and grand canonical Monte Carlo simulation
techniques. By varying the polymer-polymer interaction the state of the polymer
is changed from melt to glassy. Correspondingly the solvent density profile 
changed from error function like to a sharp front, characteristic of Case II
type transport, when Berthelot's rule is 
applied for the solvent-polymer interaction. The weight gain by the polymer 
matrix increased as $t^{1/2}$ in agreement with Fickian diffusion, even for the 
case of a glassy polymer.
This suggests that the precursor of the front is Fickian in agreement with
recent experimental observations that characterize Case II diffusion by a sharp
concentration front with Fickian type precursor.\cite{durning95,hassan99,sanopoulou01,stamatialis02} The front, however, does not
move in the time scale of our simulation. From simulation of equilibrated
solvent-polymer solution it was found that the glassy system 
with Berthelot's rule applied for the cross term is immiscible except in the 
dilute limit suggesting that the front may not move in to the polymer matrix.
Increasing the solvent-polymer interaction enhanced the solubility of the 
system without changing the nature of the diffusion process.

The solvent concentration profiles have been fitted using the one-dimensional
Fick's model of the diffusion process. The diffusivity, $D(c)$, shows strong
dependence on the state of the polymer. Far above the glass transition $D(c)$
is approximately constant and then becomes concentration dependent as the 
polymer becomes glassy. The shape of the concentration profile and the behavior
of $D(c)$ is found to be directly related. The diffusivity is constant when the
solvent concentration profile is concave, shows exponential dependence on
solvent concentration when the solvent concentration profile is convex.

The diffusivity as a function of solvent concentration was also determined
using the Darken equation for simulations of equilibrated solvent-polymer 
solution. The diffusivity calculated using this approach is in good agreement 
with the diffusivity calculated from the solvent concentration profile.

The advantage of the Darken approach for determining $D(c)$
is that it requires much smaller system sizes than for the
direct simulation of the interdiffusion process. However, each
solvent concentration c has to be determined separately and the
simulation time required to determine the $D_c(c)$ and the fugacity $f$ 
are quite long compared to the interdiffusion studies. This is
because the scaled concentration profiles superimpose even
after relatively short times. 

\section{Acknowledgments}
Sandia is a multiprogram laboratory operated by Sandia Corporation,
a Lockheed Martin company, for the United States Department of Energy's National
Nuclear Security Administration under Contract No. DE-AC04-94AL85000.







\bibliography{references}
\bibliographystyle{apsrev}

\clearpage

\end{document}